\documentclass{article}

\newcommand{\authname}[1]{{\footnotesize\sffamily\bfseries #1}}
\newcommand{\authadd}[1]{{\small\rmfamily\itshape #1}}

\usepackage{amsfonts}
\usepackage{amsmath}
\usepackage{amssymb}
\usepackage{graphicx}
\usepackage{subfig}
\usepackage{theorem}
\usepackage{xspace}
\usepackage{bbold}
\usepackage{color}
\usepackage{float}
\usepackage{epsf}
\usepackage{natbib}

\newcommand{\set}[1]{\ensuremath{\{#1\}}}
\newcommand{\formula}[1] {\begin{displaymath} \begin{array}{l} #1 \end{array} \end{displaymath} \xspace}
\newcommand{\markedformula}[2] {\begin{equation} \label{#1} \begin{array}{l} #2 \end{array} \end{equation} \xspace}
\newcommand{\Cic} {\mathcal{C}}

\theoremstyle{break}

\newtheorem{example}{Example}

\theoremstyle{plain}

\newtheorem{statement}{Statement}

\setcitestyle{authoryear,round,aysep={}}

\begin{document}

\sloppy

\title{Rational Instability in the Natural Coalition Forming}

\author{\authname{Galina Vinogradova}\footnote{galina.vino@gmail.com}\\[2pt]
\authadd{CREA - Center of Research in Applied Epistemology, Ecole
Polytechnique, 32 boulevard Victor} \\
\authadd{Paris, 75015, France}\\
\and
\authname{Serge Galam}\footnote{serge.galam@polytechnique.edu}\\[2pt]
\authadd{CREA - Center of Research in Applied Epistemology, Ecole
Polytechnique, 32 boulevard Victor} \\
\authadd{Paris, 75015, France} }

\date{}

\maketitle

\begin{abstract}

We are investigating a paradigm of instability in coalition forming
among countries, which indeed is intrinsic to any collection of
individual groups or other social aggregations.

Coalitions among countries are formed by the respective attraction
or repulsion caused by the historical bond propensities between the
countries, which produced an intricate circuit of bilateral bonds.
Contradictory associations into coalitions occur due to the
independent evolution of the bonds. Those coalitions tend to be
unstable and break down frequently.

The model extends some features of the physical theory of Spin
Glasses. Within the frame of this model, the instability is viewed
as a consequence of decentralized maximization processes searching
for the best coalition allocations. In contrast to the existing
literature, a rational instability is found to result from forecast
rationality of countries.

Using a general theoretical framework allowing to analyze the
countries' decision making in coalition forming, we feature a system
where stability can eventually be achieved as a result of the
maximization processes. We provide a formal implementation of the
maximization principles and illustrate it in the multi-thread
simulation of the coalition forming. The results shed a new light on
the prospect of searches for the best coalition allocations in the
networks of social, political or economical entities.


\end{abstract}

\newpage

\section{Introduction} \label{intro}

The subject of our study is the phenomena of instability in
coalition forming intrinsic to any collective of individual groups.
To illustrate the presentation keeping it close to topical interest,
we address the coalition forming in an aggregate of countries.
However, the discussion and the results can be applied to any type
of collective in political, social or economical sciences.

For centuries, countries have been undergoing parallel processes of
association into unities seeking mutual support and cooperation, and
fragmentation of the unities due to the bilateral conflicts.
Association and fragmentation are driven by the character of
relationships and interactions built by the countries through the
history. These interactions have formed very strong bilateral mutual
propensities between the countries, the propensities to cooperate or
to conflict.

These propensities are the key factors that determine the
association into coalitions. Like in the well known saying "the
enemy of an enemy is a friend", we assume only two competing
coalitions. Guided by the postulate of lowest energy, we expect that
positive propensities encourage countries to ally to the same
coalition, while negative ones encourage them to affiliate with the
opposing coalition. Thus, the best coalition allocation -- the one
where the countries have most comfortable and beneficial position,
is achieved in an amalgam of conflict and cooperation. Cooperation
between countries maximizes their benefit (minimizes their local
energy) in case of positive propensity. Negative propensity, on the
contrary, minimizes the benefit of cooperation, while it maximizes
the benefit of conflict.

In the model of countries collective we assume that no external
forces incite the interactions between the countries. Consequently,
the coalitions are formed only through the attraction or repulsion
caused by the historical bond propensities between the countries.

Spontaneous and independent evolution of the countries mutual
propensities produce a priori a very intricate circuit of bilateral
bonds. They are the cause of contradictory association within the
coalitions which may include conflicting countries brought together
attracted by common allies. Such associations tend to be unstable,
they break down frequently, forming new associations.

The nominal model is referred as \emph{natural model of coalition
forming}. The model resembles the Statistical Physics model of Spin
Glasses \citep{SGM}, which is an idealized model of bulk magnetism
represented by a collection of interacting spins -- atoms acting as
a tiny dipole magnet with a mixture of ferromagnetic and
anti-ferromagnetic couplings. The countries are compared to the
magnetic dipoles which interact with each other and align themselves
in order to attain the most comfortable position. The collection of
spins forms a disordered material in which the competing
interactions cause high magnetic frustration -- changes of spins
with no energy cost.

In order to reproduce more realistically a system of individual
actors that possess rationality, we suppose the countries observe
and make decision before processing to any change. This provides a
basis to a \emph{rational instability}, which is not only the
no-cost frustrations appropriate to the spins, but also the cost
frustrations driven by the future benefit of planned changes.
Rationality of the instability makes the main difference between the
Spin Glass and the present model.

Spin Glass formulations in Statistical Physics are set out in
\citep{SGM} and \citep{GT}. The modeling of complex social
situations using Statistical Physics has started from \citep{SYY}.
However, the subject of the instability and the stabilization in
coalition forming using the model is rather recent. The authors of
\citep{MG} study a model of collective decision making combining
Social Psychology hypotheses with recent concept of Statistical
Physics. Then, the coalition as a form of aggregation among a set of
actors (countries, groups, individuals) has been studied using
concepts from the theory of Spin Glasses \citep{Axel, Flo, FVS, GC,
SPC, TBISCF, APIM, SDO}.

The seminal paper \citep{Axel} applies the Spin Glass model to the
aggregation and the alignment of actors. In \citep{FVS}, this
application has been shown to be insufficient to describe the
instabilities of coalitions. The author extends the idea by
combining both random-site and random-bond Spin Glass models. He,
then proposes a political model allowing to reproduce the
incompatible interactions in the formation of coalitions. Various
social application of the model were suggested. For example,
\citep{SPC, TBISCF} use it to explain the formation of coalitions
and suggest its social and political application. The dynamical
analogue of the model and it's viability is discussed in
\citep{GAVDP}.

There is also literature that studies social or inter-agents
stabilization using a different class of models. \citep{SBN}
consider transition of social network to a balanced state by
dynamical changes of sign of the mutual links, and \citep{DSC}
feature the formation of coalition by dynamical collective decision
making.

The literature studying the coalition forming based on the Spin
Glass model analyze the phenomena within a Markovian frame of
spontaneous instability. In contrast, the current work assumes long
horizon rationality of the actors to develop the rational
instability framework of coalition forming. This allows to explore
and study the complex behaviors among individual actors possessing
rationality when they form coalitions.

Let us note that, while in Spin Glass model the Game Theory
equilibrium represents a stable configuration of the spins, the
equilibrium guarantees no stability in a model where the agents do
not stop searching upon achieving their local maximums. Therefore,
the game-theoretic tools are not appropriate for the stabilization
objective of our model.

In this work, we extend the model proposed by \citep{SPC} and
provide a detailed analysis of the instability rooted in the
coalition forming as a decentralized maximization processes driven
by each country's objective to attain its best coalition
allocations. We study the terms of existence of the optimal and
stable coalitions, and the ways to reach them in the ranges of
extensive or limited rationality. This allows to shed a new light on
the interesting and little explored phenomena such as information
manipulation and non-optimal stability, which are peculiar to any
network of selfish actors.

Among the results, the model provides an explanation of why in most
cases in practice where the rationality prevails, the instability
constantly follows the natural coalition forming.

The model's theoretical framework is supplemented by historical
examples that justify the practical use of the model. We also
provide the description and the illustration of a multi-threaded
simulation of coalition forming in the natural model that is created
for this particular aim. Finally, we provide concluding remarks and
discuss the directions of the further study.

\section{Natural Model of Coalition Forming}\label{sec_model}

How does the historical and geographical background impact the
current interactions between the countries? Based on their
historical experience the countries decide weather to agree or
disagree on the present policies. Respectively, the agreement is
unfavorable to the countries which went thought of rejection. The
same way, the disagreement is unfavorable to the countries that
passed through a conflict in their past. The primary mutual
propensities between countries are the issues of their historical
experience so that their character of cooperation or conflict can
hardly be changed. Those propensities affect all the subsequent
interactions and exchanges between the countries.

Here, we present the \emph{natural model} of coalition forming that
describes a system of countries maintaining short range interactions
which are guided by the primary mutual propensities.

Consider a group of $N$ independent countries which had experienced
geographic, cultural or economic interactions during their history.
The countries are respectively denoted by characters or indicators
ranging from $1$ to $N$. Each country thus makes its choice among
two options $+1$ and $-1$, corresponding to two possible coalitions.
The same choice allies the countries to the same coalition, while
different choices separate them into the opposite ones. According to
the postulate of minimum conflict, being part of the same coalition
benefits the countries with the propensities to cooperate, while
countries inclined to conflict bear loses from the cooperation. A
country is represented by a discrete variable that can assume one of
the two state values $S=+1$ and $S= -1$. The combination $S=
\set{S_1, S_2, S_3, \dots, S_N}$ makes up the \emph{configuration}
of the choices of the countries. The configuration of choices, as
well as it's inverse $S = \set{-S_1, -S_2, -S_3, \dots, -S_N}$ by
symmetry, represents a particular configuration of coalitions in the
system.

Consider any two countries $i$ and $j$, and denote by $J_{ij}$ the
value that measures the degree and the direction of the historical
exchanges between the countries. $J_{ij}$ represents the bond of
original propensity between the countries, which is symmetric,
$J_{ij}= -J_{ji}$, and may vary for each pair of countries. $J_{ij}
= 0$ when no mutual bond exists between the countries, which
represents an absence of a direct exchange. We shall describe by
$J_{ij} S_iS_j$ the measure of the interaction between the countries
as a function of their choices. In case the countries agree, i.e.,
when $S_i = S_j$, a positive value of $J_{ij}$ results in the
positive effect for both the countries. When $J_{ij}$ is negative,
the cooperation between the countries results in a negative effect
on both the countries. That would agree with a widespread notion
saying that conflicting countries gain from disoperation while incur
losses from cooperation.

Thus, the propensities either favor the cooperation $(J_{ij} > 0)$,
the conflict $(J_{ij} < 0)$, or signify the ignorance $(J_{ij} = 0)$
where neither the agreement nor the disagreement influences the
outcome of the countries.

\begin{figure}[!ht] \centering
\includegraphics[width=1.5in]{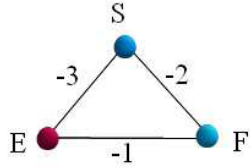} \caption
{Triangle of three countries connected by negative bond propensities
-- $ESF$ conflicting triangle.}\label{engspfr}
\end{figure}

For the sake of visualization, we represent the system of countries
as a connected weighted graph with the countries in the nodes and
the bilateral propensities as the weights of the respective edges,
(see Figure (\ref{engspfr})). We take red (dark) color for $+1$
choice and blue (light) color for $-1$ choice.

Sum of the measures of all the interactions of a country $i$ in the
system is the net gain of the country
\markedformula{countr_gain}{H_i(S) = S_i\sum_{j\ne i}J_{ij}S_j}. The
total gain in the system in configuration $S$ is measured by the
total of the contributions in the configuration
\markedformula{syst_gain}{\mathcal{H(S)} = \frac{1}{2}
\sum_i{H_i(S)}.}

\begin{figure}[!ht] \centering
\includegraphics[width=2.3in]{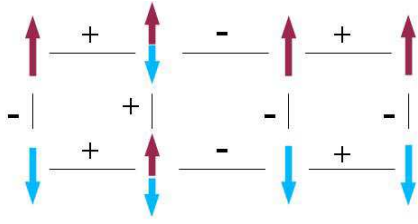} \caption
{Ising model of $8$-spins with negative pair interactions. Spins
that shift their states with no loss of energy, called frustrated
spins, are marked by both up and down the states. A Spin Glass in an
unstable disorder.}\label{spin_glass}
\end{figure}

The natural model is formally identical to the Ising Model of Spin
Glass in Statistical Physics. The model consists of $N$ discrete
variables of magnetism $\set{S_i}_{1}^{N}$, called spins, that can
be in one of two states \emph{up} or \emph{down}. See figure
(\ref{spin_glass}). The spins, arranged in a lattice or a graph
interact at most with its nearest neighbors. Spins with the same
states are associated to each other, and disassociated from those
with the opposite states.

In this analogy, the total gain $\mathcal{H(S)}$ in the system is
identical to the Hamiltonian of an Ising random bond magnetic
system. The model of the Spin Glass has been solved numerically with
much effort (\citep{SGM}).

\subsection{Illustrating of The Natural Model Trough an Historical Example}

As an example of an unstable system we suggest to consider the
typical conflicting triangle of Spain, England and France, which
were alternatively enemies and allies during a long period of time.
For these countries, the period of $1521$ -- $1604$ has been marked
by the series of land and sea wars driven by the historical
background of colonization, religion and by the naval technical
progress. The commercial rivalry between England and Spain, and
political and religious ambitions of France were the major forces
that pushed Europe into wars. The conflicts were usually initiated
between any two of the countries with the third one joining the one
or the other side. Accordingly, the historical background of the
countries during this period has defined the particular distribution
of mutual negative propensities.

The natural model is shown schematically in the $ESF$ conflicting
triangle in Figure (\ref{engspfr}), where abbreviations $S, E$ and
$F$ stand for the countries' names. The choices of the countries
$S$, $E$ and $F$ are represented by the state variables $S_E$,$S_S $
and $S_F$ respectively. Different colors attached to the state nodes
of the countries correspond to different coalitions the countries.
The state nodes are linked by the following original propensities
$J_{SE} = -3$, $J_{SF} =-2$, $J_{EF} = -1$. The value of
propensities are illustrative only and this is their relative
magnitudes are of importance: the conflict between Spain and France
is less deep than the one between Spain and England, yet it is
deeper than the one between England and France.

\section{Instability in Coalition Forming}

During the coalition forming, at any particular configuration, a
country may observe that another configuration exists where it can
reach a higher gain. When it happens, the country shall take
appropriate changes in order to take advantage of this opportunity.
The sequence of the changes constitutes the process of maximization
of the countries' gains.

Maximization, which is a search for the most beneficial coalitions
setting is found to produce an instability. When the countries
eventually attain a common satisfactory configuration, the
instability is temporary. When it is a permanent one, the
maximization of the countries' gains is an endless process.

\subsection{Geometric Terms of Instability}

Let us recall that in the natural model we extend the notion of
instability of the physical systems where it is only caused from
frustration of spins due to an immediate improvement or due to no
cost of changing. In the broader sense, frustrations due to intended
changes aimed to improve the current gains in further steps are also
included. Such improvements are peculiar to the system of countries
which are able to anticipate and adapt to the changes of others.
Spins, unlike countries, are able to evaluate only the effect from
the immediate changes. In order to be able to predict such an
unstable character of a system of countries, we provide the
necessary and sufficient conditions of the instability.

Let us take a close look at the heart of the instability. Consider
separately from the rest of the system, two countries $i$ and $j$
connected by their propensity bond $p_{ij}$. When $p_{ij}$ is the
only link between the countries, we can easily valuate the
contribution from cooperation or conflict between the countries to
their gain: $p_{ij}S_iS_j$. What complicates the evaluation of their
alternative choices is the presence of the other countries
connecting between $i$ and $j$ indirectly.

\begin{figure}[!ht]
\centering
\includegraphics[width=1.3in]{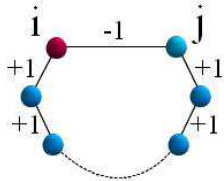} \caption
{Circle connecting between the countries $i$ and $j$ by a negative
and a positive links at the same time produces an everlasting
competition between the countries.}\label{circ_ex}
\end{figure}

By a circle in a graph of countries we understand a selection of
nodes linked into a closed path by the propensity bonds. Imagine a
situation where the countries $i$ and $j$ is part of a circle and
assume without loss of generality that their mutual propensity is
negative while the propensities all along the rest of the circle are
positive (see Figure (\ref{circ_ex})). In this case, both $i$ and
$j$ maximize their gain in conflict with the other and in
cooperation with the rest countries on the circle. Such an
arrangement creates an everlasting competition between $i$ and $j$
for this maximizing arrangement. The countries continuously shift
their choices that produces the instability.

We are now ready to identify the instability formally. Denote a
circle of countries by $\Cic$ and the countries composing the circle
by $1,2, \ldots, k$. Then the characteristics of the phenomena of
instability in Spin Glasses \citep{GT} can be interpreted in the
terms of the natural model: \markedformula{stab}{$\emph{if there is
a closed circle of countries on which the product of total
propensities is negative},$\\ \hspace{2in} \Pi_{i,j \in \Cic} p_{ij}
< 0 \hspace{0.02in}\\ $\emph{then the system is unstable}.$}

Terms (\ref{stab}) feature the instability in a system of countries
as a direct consequence of geometry of the system -- presence of a
negative circle. In order to identify the instability, the terms
require to examine all the possible circles in the system, while
provide no countries' competition data. We call (\ref{stab})
geometric terms of instability.

\begin{figure}[!ht]
\centering
\includegraphics[width=1.75in]{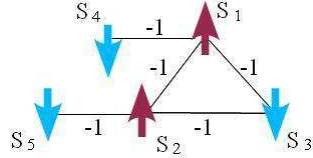} \caption
{Configuration which is stable in Spin Glass and unstable in the
system of countries.}\label{limt_instab}
\end{figure}

The geometric instability terms, while originate from physics,
provide only the necessary condition for the instability in the Spin
Glass. The Spin glass with a negative circle can be stable when, for
example, an interaction with a neighbor spin shifts the energy of
the spin form zero to negative and thus keeps the spin from
fluctuating. This case illustrated in Figure (\ref{limt_instab}).

In the extended sense of instability, (\ref{stab}) provides also a
sufficient condition of instability, the one that induces the
endless competition among the countries for beneficial coalitions.

\subsection{Optimal Configurations and Analytical Terms of Instability}

By making a choice, a country $i$ realizes one of two possible
system's configurations $S^{+i}$ and $S^{-i}$ which differ by the
country $i$'s choice $S_i$. In order to select the preferable one,
the country needs to classify the configurations by the order of
preference based on the potential gain.

A preference order over a set of elements in general, is defined by
a binary relation on the set which is a collection of ordered pairs
of its elements. Each country $i$'s function of gain $H_i$ sets the
binary relation of preference $\le_{i}$ over the set of all system's
configurations as follows:  configuration $S'$ is less preferred
than $S''$, ~{$S' \le_{i} S''$} if ~{$H_i(S') \le H_i(S'')$}. The
total gain in the system $\mathcal{H}$, in turn, sets the relation
$\le_{sys}$ where ~{$S' \le_{sys} S''$} if ~{$\mathcal{H}(S') \le
\mathcal{H}(S'')$}. Each of the relations $\set{\le_{i}}_{i=1}^N$
and $\le_{sys}$ defines a partial order over the set of
configurations -- the order that indicates a relation between some
elements while others remain unrelated. For each particular partial
order, those unrelated configurations yield the same gain to the
country and therefore belong to the same equivalence classes. The
equivalence classes thus are strictly ordered from minimal to
maximal.

Each equivalence class, unifying under the same gain the
configuration and its inverse, contains at least two configurations
which a priori leaves only  half of the possibilities of different
gains, $2^{N-1}$.

The maximal configurations of a country are contained in its maximal
equivalence class. Similarly, the system's maximal configurations
are in the system's maximal equivalence class. Consequently, the
common maximal configurations, which are those that satisfy all the
countries, lie in the intersection of the maximal equivalence
classes of all the countries. If we denote by $C_i^{\gamma_i}$ an
equivalence class of country $i$ corresponding to gain $\gamma_i$,
and by $C_i^{\Gamma_i}$ the maximal equivalence class, then the set
of common maximal configurations is determined by
\markedformula{max_conf}{\mathcal{S}^{\Gamma_i} = \cap_{i=1}^{N}
C_i^{\Gamma_i}.}

If $\mathcal{S}^{\Gamma_i} \ne \emptyset$, then the countries share
the same maximal configuration in the system. The configuration is
an \emph{optimal configuration}, the one that satisfies all the
countries and guarantees stability of the system.

\begin {statement}Formula (\ref{max_conf}) provides new terms of the instability
reading that if\\
\markedformula{instab_qual}{\hspace{0.05in} \cap_{i=1}^{N}
C_i^{\Gamma_i} = \emptyset} than the system is not stable in any
configuration.
\end{statement}
This is the analytical terms that, along with the indication of
system's stability, provide particular optimal configurations, as
well as the respective values of gains of the countries.

Here, the phenomena of instability is explained by the fact that, a
priori, preference orders of different countries do not coincide,
and consequently, maximal coalitions for one country is not maximal
for the others.

\begin {statement}Let us remark that when an optimal configuration
exists, it is unique up to inverse. \end{statement}

Assume, by contradiction, that there are two different optimal
configurations which however are not the inverse of each other,
$\mathcal{O}1$ and $\mathcal{O}2$. Then, there is at least one
country whose states differ in the two configurations, and another
country whose states agree in the configurations. As far as we
consider a connected system, there are necessarily two such
countries which are connected by a propensity bond. Denote them by
$i$ and $j$. Then, $S_{i}^{O1} = - S_{i}^{O2}$ and $S_{j}^{O1} =
S_{j}^{O2}$.

Assume, without loss of generality, that $J_{i,j} = J > 0$. Since
$\mathcal{O}1$ and $\mathcal{O}2$ are both optimal then
$\mathcal{H}_i(\mathcal{O}1)= \mathcal{H}_i(\mathcal{O}2)$. Since
they differ by the states of $i$, there is country $k$ such that
$J_{i,k} = -J$ and the following holds:
$\mathcal{H}_i(\mathcal{O}1)= \mathcal{H}_i(\mathcal{O}2) - J_{i,j}
+ J_{i,k}$ and $\mathcal{H}_i(\mathcal{O}2)=
\mathcal{H}_i(\mathcal{O}1) + J_{i,j} - J_{i,k}$. Then, either there
is a negative circle connecting $i,j$ and $k$, or there is another
configuration $\mathcal{O}3$ such that $\mathcal{H}_i(\mathcal{O}3)
= \mathcal{H}_i(\mathcal{O}1,\mathcal{O}2) + J_{i,j} + J_{i,k}$.
Which is a contradiction to the maximality of $\mathcal{O}1$ and
$\mathcal{O}2$.

\subsection{Illustration of Stable and Unstable Systems}

Consider the example having no optimal configuration.
\begin{example}[Unstable System]\label{ex_cong_trian}
Consider again the traditional conflicting triangle of Spain,
England and France as in Figure (\ref{engspfr}). We feature the
instability of the triangle from the perspective of the countries
preference orders over the configurations.

Each configuration is of the form $S = (S_E, S_S, S_F)$. The $8$
different configurations in total represent $4$ different
coalitions. The functions of the countries' gains which are $H_E =
-3S_E S_S -S_E S_F$, $ H_S = -3S_S S_E -2S_S S_F$, and $H_F = -S_F
S_E -2S_F S_S $
produce the following equivalence classes of the countries: \\

The equivalence classes of $E$:
\formula{ \Gamma_E = 4 \hspace{0.1in}: {C_E^{4}= \set{(-1, +1, +1), (+1, -1,-1)},}\\
\Gamma_E = 2  \hspace{0.1in}: {C_E^2 = \set{ (+1, -1,+1), (-1, +1, -1) },}\\
\Gamma_E = -2 \hspace{0.1in}:{C_E^{-2}= \set{(+1,+1, -1), (-1,-1, +1)},}\\
\Gamma_E=-4 \hspace{0.1in}:{C_E^{-4}=\set{(+1,+1, +1),(-1,-1,-1)}.}}

The equivalence classes of $S$:%
\formula{ \Gamma_S = 5 \hspace{0.1in}: {C_S^{5}= \set{(-1, +1, -1), (+1, -1,+1)},}\\
\Gamma_S=1 \hspace{0.1in}: {C_S^1 = \set{ (+1, -1, -1), (-1, +1,+1) },}\\
\Gamma_S=-1 \hspace{0.1in}:{C_S^{-1}= \set{(+1,+1, -1), (-1,-1, +1)},}\\
\Gamma_S=-5 \hspace{0.1in}:{C_S^{-5}=\set{(+1,+1, +1),(-1,-1,-1)}.}}

The equivalence classes of $F$:%
\formula{ \Gamma_F = 3 \hspace{0.1in}: {C_F^{3}= \set{(-1, -1, +1), (+1, +1,-1)},}\\
\Gamma_F=1 \hspace{0.1in}: {C_F^1 = \set{ (+1, -1, +1), (-1, +1,-1) },}\\
\Gamma_F=-1 \hspace{0.1in}:{C_F^{-1}= \set{(-1,+1, +1), (+1,-1, -1)},}\\
\Gamma_F=-3 \hspace{0.1in}:{C_F^{-3}=\set{(+1,+1, +1),(-1,-1,-1)}.}}

As we can see, there is no common maximal configuration in the $ESF$
triangle since the intersection of the maximal equivalence classes
of the three countries is $C_E^{4} \cap C_S^{5} \cap C_F^{3}=
\emptyset$.

The system thus shifts between the individual maximal coalitions
infinitely, driven by each country's endless search for its best
position.

Figure (\ref{conf_loop}) shows series of transitions in the triangle
that cycles exhibiting the endless instability.
\begin{figure}[!ht]
\centering
\includegraphics[width=4.2in]{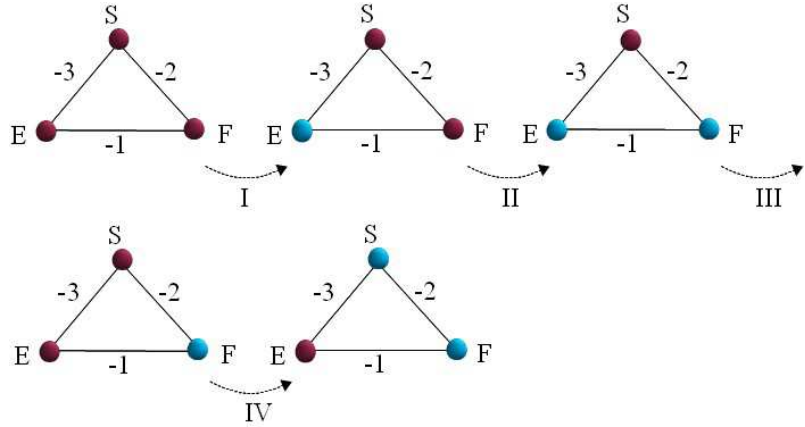} \caption
{The diagram of successive configurations transition in the $ESF$
conflicting triangle.}\label{conf_loop}
\end{figure}

The progression starts from $S=(+1,+1, +1)$, which is the worst
configuration to all the countries. At the first step, $E$ changes
its state to $-1$ which yields the country an immediate benefit of
$\Gamma_E = 4$. This change shifts the countries to the equivalence
classes $C_E^{4}$, $C_S^{1}$ and $C_F^{-1}$, respectively. At the
second step, country $F$ make change and moves to $C_F^1$
equivalence class. Then, at the step $III$, $E$ makes a change aimed
to get back its best configuration in some further step. Finally, at
the step $IV$, country $S$ changes to $-1$, which brings the
countries back to the original equivalence classes $C_E^{4}$,
$C_S^{1}$ and $C_F^{-1}$.
\end{example}

Let us now turn to the example of a system where an optimal
configuration exists.
\begin{example}[Stable System]

A snap shot of the short period in the midst of war against Spain,
when England and France were favorable to each others, illustrates a
system where an optimal configuration exists (see Figure
(\ref{engspfr_st})).

\begin{figure}[!ht] \centering
\includegraphics[width=1.5in]{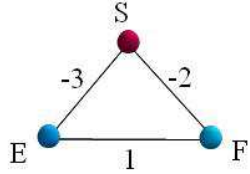} \caption
{In this arrangement, the $ESF$ triangle has an optimal
configuration of coalitions -- $\set{S}$, $\set{E,F}$, and thus is a
stable system.}\label{engspfr_st}
\end{figure}

The maximal possible gains of $E$, $S$ and $F$ are $H_E = 4$, $H_S =
5$ and $H_F = 3$, respectively, and the maximal equivalence classes
of the countries coincide: $C_E^{4}= C_S^{5}= C_F^{3}= \set{(+1, -1,
+1), (-1, +1, -1)}$. Consequently, there are two optimal
configurations in the system, $~{C_E^{4} \cap C_S^{5} \cap C_3^{2}
\cap C_F^{3} = \set{(+1, -1, +1), (-1, +1, -1)}}$, each of which
guarantees its stability.
\end{example}

\section{Rational Maximization of Countries}

Maximization of countries' gains is a rational process which can be
defined in a general theoretical framework linking between the
system's configurations information and the law of the configuration
dynamics.

\subsection{Maximization As a Sequence of Individual Choices}

As we have mentioned, maximization is a sequence of the changes made
by the countries that aim to improve their benefits. In our model,
where the countries are individual actors possessing rationality, a
change is made by the rational components such as the observation
and the decision making.

Let us examine the mechanisms that stand behind the changes. At any
configuration $S$, a country $i$ can make a choice for its state.
The alternative is either to keep the current state, or to invert it
and thus to move to configuration $S^{-i}$. Making the choice
consists of two successive phases -- observations and decision
making. The phase of observation involves featuring the equivalence
classes and looking for transitions between the configurations
aiming at the immediate or the further improvements of the gain. The
phase of the decision making, in turn, involves the selection of an
immediate reply based on the results of the observation phase.
Denote by $O_i(S)$ the set of desirable configurations and by
$D_i(S)$ the immediate reply, $D_i(S)\in \set{S, S^{-i}}$.

The decision making phase can be based either on the \emph{best
reply} principle, which accepts the choices that improve the gain
immediately, or on the \emph{forecast} principle, which takes into
account the choices that improve the gain in a further step. The
choice between the equivalent alternatives can be based either on
the principle of \emph{random move} or on the principle of
\emph{anticipation of feedback} from the others countries.

The following example illustrates observation and decision making by
the countries of the $ESF$ conflicting triangle in the series of
transition of configurations.

\begin{example} Figure (\ref{conf_move}) shows several possible
configuration transitions in the conflicting triangle as the result
of particular combinations of observations and decisions made by the
countries in the $ESF$ triangle.

\begin{figure}[!ht] \centering
\includegraphics[width=5.2in]{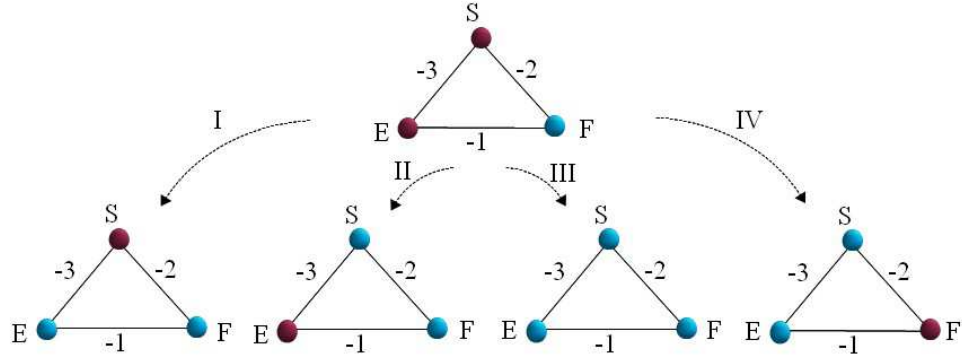} \caption
{The diagram of transitions of configurations prompted by the
decisions in the $ESF$ conflicting triangle.}\label{conf_move}
\end{figure}

The initial configuration $S = (+1, +1, -1)$ belongs to the
equivalence classes $C_E^{-2}, C_S^{-1}$ and $ C_F^{3}$ of $E, S$
and $F$, respectively. Let us analyze the $I-IV$ transitions shown
in the Figure (\ref{conf_move}).
\begin{itemize}
\item[$T_{I}$:]{ Observation by
$E$ results in $O_E(S) = \set{(+1, +1, -1), (-1, +1, -1)}$ where
configuration $(-1, +1, -1)$ is for the immediate improvement and
$(+1, +1, -1)$ is for the improvement in a further step, which will
eventually lead to the $E$'s maximal configuration $(+1, -1, -1)$.
$E$ makes the decision $D_E(S) = (-1, +1, -1)$ and thus moves the
system to the equivalence classes $C_E^2, C_S^{5}$ and $C_F^1$.

Note that, while the change satisfies only country $S$, it is
positive for the system as a whole; the new configuration is the
system's maximum.}

\item[$T_{II}$:]{ If the change is done by
country $S$ who makes the decision $D_S(S) = (+1, -1, -1)$, the new
configuration belongs to the classes $C_E^{4}, C_S^1$ and
$C_F^{-1}$.

Though the configuration improves the gain of the system itself, it
is not the system's maximum.}

\item[$T_{III}$:]{If the change is performed by both the countries $E$ and $S$ simultaneously,
with $D_{E}(S) = (-1, +1, -1)$ and $D_{S}(S) = (+1, -1, -1)$, the
resulting configuration is $S = (-1, -1, -1)$ which belongs to the
equivalence classes $C_E^{-4}, C_S^{-5}, C_F^{-1}$.

This is the worst case for each of the countries, as well as for the
system itself.}

\item[$T_{IV}$:]{When the changes are performed by all the countries $E$, $S$ and $F$
simultaneously, and the resulting configuration $(-1, -1, +1)$
belongs to the initial equivalence classes $C_E^{-2}, C_S^{-1} and
C_F^{3}$ .}
\end{itemize}
\end{example}

It is worth to emphasize that since each country at any time can
decide to update or not to update its state, the transitions are all
equiprobable. All the transitions are codified in the configurations
information trees described below.

\subsection{Configurations Information Tree}

From the point of view of a given country, the choices made by the
others are the direct or indirect outcomes of the country's own
decision. Such dependencies between the country's choice and the
decision of the others, while developing staring from the initial
configuration, define a tree-like structure.

We define the \emph{configurations information tree} as the
centralized structure containing all the possible system's
configurations and all the possible transitions between them. The
tree of a country $i$ is built as follows. At the root there is the
initial configuration $S$ of the system. The root splits into two
\emph{choice nodes} corresponding to the two possible choices of the
country $~{+S_{i}}$ and $~{-S_{i}}$. Each of the nodes splits into
$2^{N-1}$ \emph{configuration nodes}, which are the possible
decisions by the other countries. Starting from the configuration
nodes, the tree develops the same way as it does from the root. The
choice nodes of the tree continue to develop until they correspond
to a configuration that has already appeared in a configuration node
of the tree.

The information tree contains the transitions of any feasible
maximization processes prompted by the country. It allows the
country to make unlimited observations and to build the best reply
strategy starting from any configuration. Let us look at the
information tree of England in the $ESF$ conflicting triangle of
Figure (\ref{engspfr}).

\begin{example}[Information Tree of England in the $ESF$ Conflicting Triangle]

Figure (\ref{conf_tree}) shows the information tree of $E$. The
configuration nodes are marked by red (dark) color until they or
their inverses reappear in the tree, then they are marked by green
(light) color. The values of corresponding countries' gains are
shown next to each configuration node. The choice nodes are marked
by light blue.

The lines proceeding from a configuration node correspond to the
choices made by $E$ and the dashed lines drawn from the choice nodes
represent the transitions initiated by the others.

\begin{figure}[!ht] \centering
\includegraphics[width=6in]{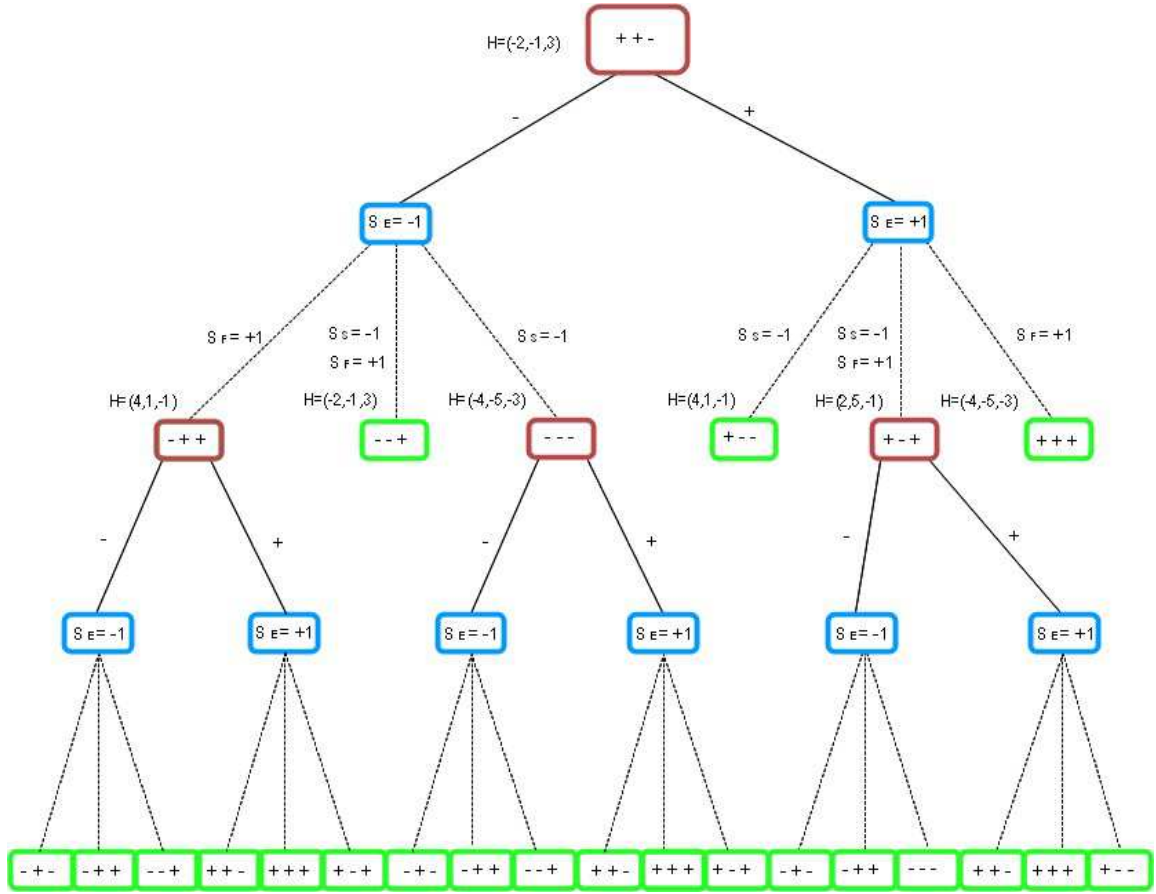} \caption
{Configurations Information tree of England in the $ESF$ conflicting
triangle. In the tree, the configuration nodes are alternated with
the choice nodes followed by the decision of the country. The root
represents the initial configuration.}\label{conf_tree}
\end{figure}

\end{example}

Observation function includes a search for the pathes in the tree
that brings to a best configuration. The size of the tree defines
the lower bound on the complexity of the search, which is a lower
bounds on the minimum amount of time required by the most efficient
search algorithm. Since only one among any two symmetric
configurations is developed, and each of the respective
configuration nodes is followed by the two choice nodes, the tree
contains $\mathcal{O}(2^{N})$ nodes. It follows that the rational
functions get harder exponentially as the system grows.

\section{A Formal Implementation of Rational Maximization}

A system of countries represents a decentralized multi-agent system
of individual actors separated by geographical and cultural
distances. The geographical location and the initial conditions
foreordain different capacities and different degrees of impact by
technological development. Those distances and differences result in
diverse levels of exposure to the system information, as well as in
divers technical capacities of observation. The differences become
more distinct with the increase of the system's size.

The complete system information available to a country and its
extensive technical capacities contrast with the limited system
knowledge and limited capacity. The limited knowledge of a country
prevents from proper observation of the other's states and
decisions. The limited capacity restricts its ability to build the
system information and to track the maximization processes. Both
limitations perturb the observation function of the country
introducing limitations in their rationality.

Another restriction in this concern is the latency in the decision
making which affects the quality of the decision output. Here, we
ignore the decision latency as an issue of a different subject.

The following cases can be distinguished in the ranges of the
classification of capacities.

\begin{itemize}
\item[$I$:]{ 
Imagine a system which has an optimal coalitions and where the
countries have extensive technical capacities. The maximization
process in the system is in fact the optimization. A correct
information, the accurate configurations data guarantee the
appropriate observation and decision making and thus are crucial for
the optimization.

The existence of unique optimum combined with the countries'
extensive technical capacities guarantees a rapid stabilization with
hight probability. However, systems that stabilize spontaneously are
rare. The probability that the system has an optimum vanish
exponentially with the size.}

\item[$II$:]{ 
The system with no optimum is more likely. This case, combined with
the rational instability, produces the infinite competition for
beneficial coalitions that trap the maximization process into an
infinite cycle. }

\item[$III$:]{
Finally, the most real situation is the system with no optimum where
some of the countries have limited system knowledge and a limited
capacity. The rationality limitations influence the maximization
which either involves an infinite competition or becomes a finite
stabilization. The latter lead to a stable state which though maybe
temporary. This case is investigated in the next section.}
\end{itemize}

\subsection{Rationality Limitations and Information Manipulation}

In order to attain the maximal benefit with high probability a
country have to be able to forecast the behavior of the others. The
forecast principle consists of screening the possible configuration
transitions that can eventually lead to the country's maximal gain.

It is not enough for a country to be only aware of its neighbors. In
order to complete a proper forecast, the country needs to keep in
view all the dependencies and bonds of the other countries linked to
it indirectly. In practice, the forecast screening bases on the
logic of the configurations information tree. Therefore,
completeness of the information tree of the country defines the
completeness of its forecasts ability.

As a result of the limited system knowledge and the limited
capacity, the country construct an incomplete. However, it is not
only the limited rationality that confines the forecast abilities.
The configuration information trees grow exponentially following the
system's size growth. As a result, a complete tree gets harder to
build, so harder it gets to forecast.

The countries with a limited system knowledge and a limited capacity
are unable to build the complete information trees. Therefore, they
are unable to perform a proper forecast of the possible transitions.
As a result, the limited system knowledge and the limited capacity
produce such rationality limitations that, even with a good will,
the countries tend to make wrong choices. The countries may neither
know if there is a common maximum, nor be able to forecast the
common behavior and make the corresponding choices.

In practice, in cases of the limited rationality, the rationality
itself ceases to be the criterium of choices and is substituted by
inspirations of all kind, such as religious, moral or cultural
codes.

Therefore, two coexisting types of countries can be distinguished in
such systems: $1)$ countries possessing the complete system
information and extensive capacities, $2)$ countries who has a
limited system knowledge and a limited capacity.

While the knowledgeable countries forecast and make choices that
should benefit them in a future step, the limited countries either
make the changes bearing immediate profit or follow the inspirations
and the cultural codes. Not having knowledge necessary to conclude
on existence of a common maximum and not being able to forecast the
future steps, the secondary actors are interested to stay within the
set of their local maximums. The countries adhere to the policy
where they undertake changes only upon an immediate improvement.

It is reasonable to assume that inspirations and cultural codes
playing a centralizing role on the limited countries can be imposed
by the knowledgeable ones that aims to achieve their benefits. Such
an influence is the essence of information manipulation phenomena.

On this basis, we refer the countries possessing the complete system
information and extensive capacities as primary actors, and the
countries who has a limited system knowledge and a limited capacity
as secondary actors.

The information manipulation can be described as acting to provide
wrong or partial information while pursuing certain objectives. The
information manipulation can be viewed as a natural consequence of
the extended rationality of the primary actors.

The combination of knowledge and ignorance may bring to
stabilization of the systems despite of the negative circles. Below
we illustrate the system that stabilizes regardless of lack of the
optimum.

\begin{example}[Information Manipulation and Non-optimal Stability]
Here, we present a system of four countries (see figure
(\ref{countr_manip})) forming two triangles, a stable and an
unstable, where the latest is identical to the $ESF$ conflicting
triangle.

Assume that country $1$ is a primary actor who possesses the
complete system information and the extensive capacities, while all
the others are the secondary actors having limited system knowledge
and limited capacities. The primary actor's node is emphasized by
the additional circle around the nodes.

The initial configuration is only profitable to country $3$.
Therefore, country $1$ initiates a maximization process by first
inverting its state to $+1$ (step I), and then convincing $2$ to
invert it's state too asserting that $3$ going to switch to $S_3 =
-1$ (step II). As a result, country $3$ changes to $S_3 = -1$
seeking an immediate improvement. The resulting configuration is the
maximal one for $1$ and is the stable one long country $3$ is unable
to lead a profitable maximization process.

\begin{figure}[!ht] \centering
\includegraphics[width=4.5in]{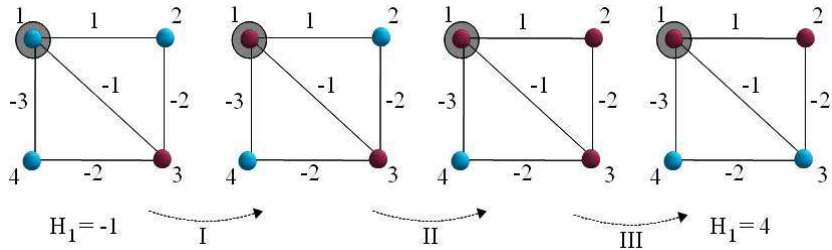} \caption
{Stabilization of a system having a negative circle by the
information manipulation.}\label{countr_manip}
\end{figure}

\end{example}

The unlimited rationality give to the primary actor (or to group of
leading countries with their own common maximum) an advantage over
the others to govern changes leading to profitable coalitions. The
widespread belief that keeping the information is secret benefits
country is fully justified in this case.

However, as soon as a secondary actor supplements the lack of the
information with the data gained by tracking the maximization
history or from the information exchange, the non-optimal stability
breaks. Therefore, a non-optimal stable state is rather a temporary
one, as it is seen in history.

The picture changes when an optimum exists. Once the existence is
known, the complete information is the common interest of the
countries. In this case, the contrary to the widespread belief about
advantage of keeping information secret is valid: disclosing
information helps the countries to achieve the common profitable
coalitions.

The phenomena of the information manipulation is intrinsic to any
system of individual actors where the respective capacities vary. As
we have noted, the frame of the natural model is applicable to the
large spectrum of social, political or economics domains, from the
dynamics of social opinion to the dynamics of interactions between
publics and private banks.

\subsection{Simulation of Coalition Forming in a Finite-size System}

To illustrate the coalition forming, we created a graphical computer
simulation of the natural model. The simulation is based on the
structure of the model, where the choice making process consists of
phases of observation and decision, as defined earlier. The
countries are represented by the simultaneously running threads
whose action readiness grows with the countries rationality. There
is no predefined order on the countries' decision making.

For the sake of simplification in simulation of individual actors,
country's observation function, instead of searching for the best
paths, is reduced to determination of state bring an immediate gain.
Then, with regards to the type of the actor, the state or its
inverse is adopted by the actor. The primary actors make a
disadvantageous change aimed put the system out of its local maximum
and to improve the gains in further steps. Such strategy enables us
to fully simulate a maximization process by following randomly the
equiprobable chains of configuration transitions.

In the simulation, the primary actors randomly make the choices due
to an expected gain. The secondary actors undertake only changes
bearing an immediate gain. They adhere to the anticipation of a
negative result that provides a benefit without jeopardizing the
whole system. This implies that no rational instability is triggered
by the secondary actors, as well as the no-cost changes.

\begin{figure}[H] \centering
\includegraphics[width=2.5in]{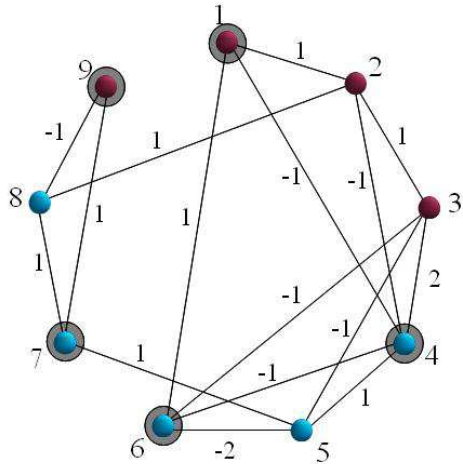}
\caption{The initial state of a system of $9$ countries. Countries
countries $1$, $4$, $6$, $7$ and $9$ are the primary actors, and
$2$, $3$, $5$ and $8$ are the secondary actors. Two groups with
disconnected primary actors are formed in the system.}
\label{simul_9_count}
\end{figure}

Here, we present the simulation of coalition forming in a
finite-size system that involves both the primary and the secondary
actors. The system in its initial state is shown in figure
(\ref{simul_9_count}).

Primary actors are countries $1$, $4$, $6$ and $7$, $9$. The
countries are emphasized by the additional circles around the nodes
in the figure. The remaining countries are the secondary actors.

The primary actors countries form two groups $\set{1,4,6}$ and
$\set{7,9}$ which have no direct connection between them.

Both groups have the local common maximums:
\formula{\mathcal{S}_{1,4,6} = \set{(1,-1,1), (-1,1,-1)} $ and
$\mathcal{S}_{7,9} = \set{(1,1), (-1,-1)}.}

The negative circles of the system are $\set{2,3,4}$,
$\set{1,2,3,4}$, $\set{3,4,5}$, $\set{3,4,6}$, $\set{1,2,3,4,5,6}$
and $\set{7,8,9}$. As it can be observed, the system consists of two
parts, each of which formes a negative circle, the right part
$\set{1,2,3,4,5,6}$ and the left part $\set{7,8,9}$.

\begin{figure}[H] \centering
\includegraphics[width=5.9in]{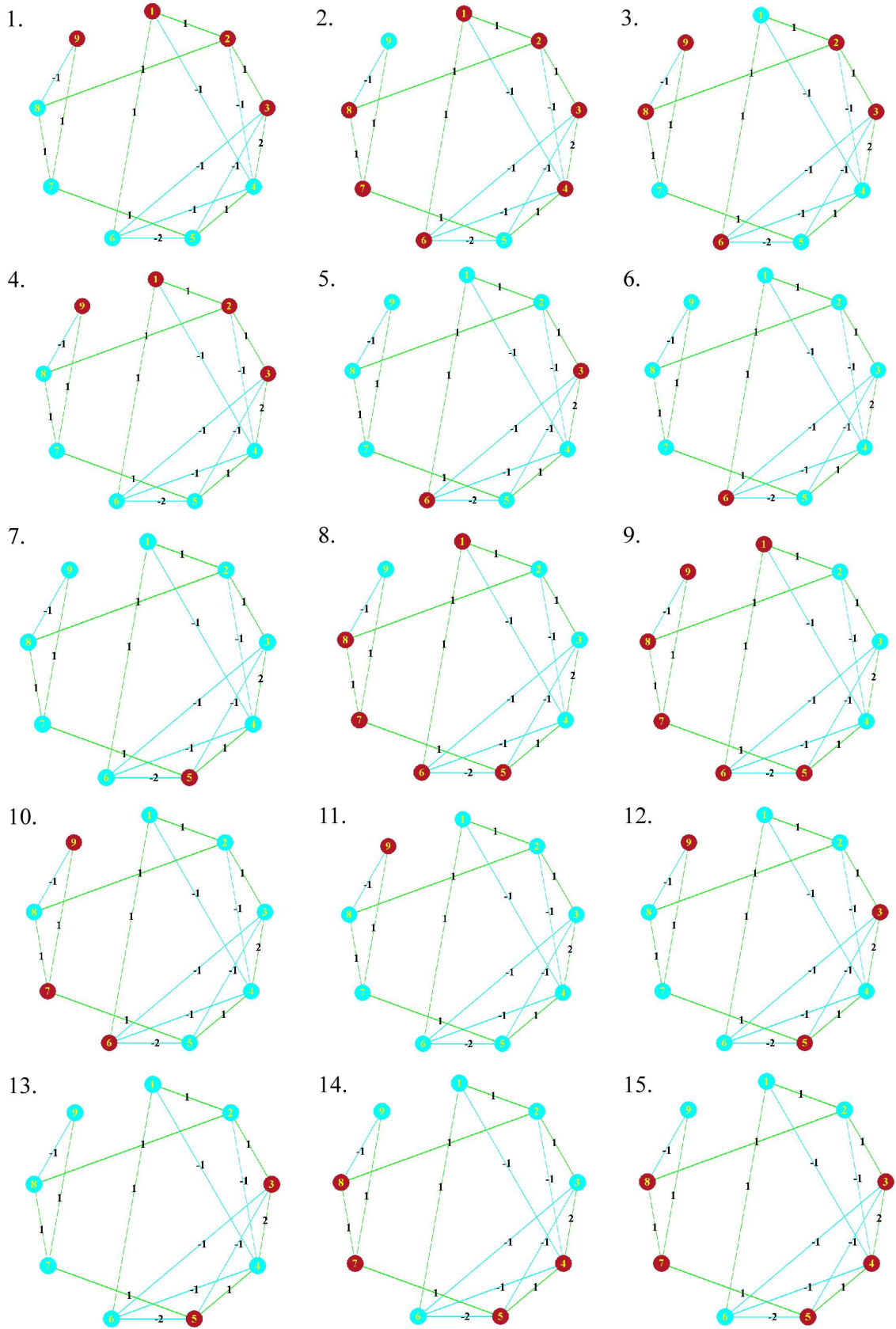}
\caption{Simulation of the coalition forming -- $15$ steps of
stabilization of the right part.} \label{Stabilization_1}
\end{figure}

Despite the fact that the right part contains several negative
circles, it is being stabilized due to particular interactions
between primary and secondary actors (see the stabilization in
Figure (\ref{Stabilization_1})). The right part of the system is
stable in the last configuration $15$ of the figure: the group of
primary actors ${1,4,6}$ meet their common maximum, and none of the
secondary actors' changes can give an immediate improvement.

Let us look at the stabilization in details:
\begin{enumerate}
\item{The initial state.}
\item{ The principal actors $4$ and $6$ simultaneously make an immediate improvement of the gain
time, the principal actors $7$ and $9$ make the change for an
expected gain, and $8$ improves its gain in reply.}
\item{ The principal actor $1$ changes for a future gain and $4$ improves its gain, $7$ and $9$ repeats the
previous change. }
\item{ Country $1$ improves its gain, $6$ makes a change for a future gain and $8$ improves its gain.}
\item{ Country $6$ improves its gain, $1$ makes a simultaneous change for an
expected gain, $2$ improves its gain in reply.}
\item{Country $3$ improves its gain in result.}
\item{Country $6$ makes a change for an expected gain and $5$ improves
its gain in reply.}
\item{Country $1$ makes a change for an expected gain and $6$ improves
as result, $7$ makes the change for a future gain and $8$ improves
in result.}
\item{ Country $9$ makes the change for an expected gain.}
\item{Countries $1$, $5$ and $8$ improve their gains.}
\item{ Countries $6$ and $7$ make the change for their expected
gains.}
\item{Countries $5$ improves its gain.}
\item{Country $9$ makes the change for a future gain.}
\item{Countries $3$ and $4$ improve their gains, $7$ makes the change
for a future gain and $8$ improves as result.}
\item{Countries $3$ improves its gain.}
\end{enumerate}

\begin{figure}[H] \centering
\includegraphics[width=5.9in]{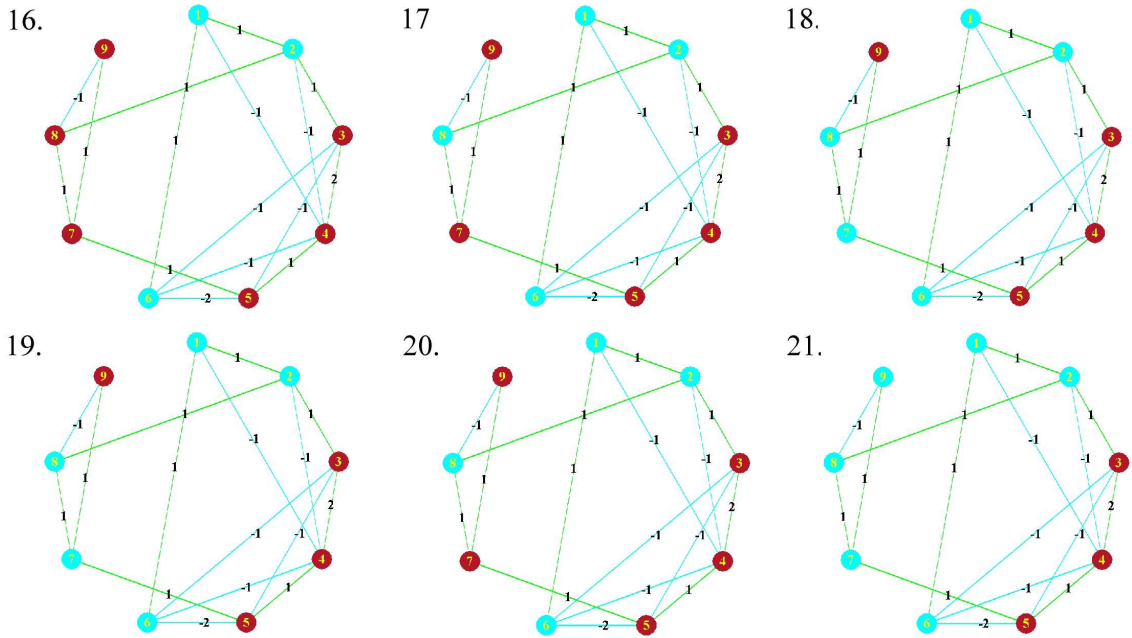}
\caption{Partial instability -- the instability of the left part
versus the stability of the right part. } \label{Stabilization_2}
\end{figure}

In the given disposition, the stability of the right part is solid
enough not to be broken despite of the fluctuation in the left part.
Figure (\ref{Stabilization_2}) ) shows the instability of the
triangle $\set{7,8,9}$, from the step $16$ to the step $22$.

However, the stability of $\set{1,2,3,4,5,6}$ is fragile since it
breaks even upon a slight change of a propensity between the
secondary actors. For instance, if the propensity between the
countries $2$ and $8$ rises to $2$, the instability of the left part
is propagated to the stable right part. Another example is when the
propensity between countries $3$ and $5$ decreases to $-3$. Then,
country $3$ will be always able to improve its gain immediately from
$-1$ to $1$ by inverting its choice.

\section{Conclusions and Remarks}

The historical examples such as the conflicting triangle of England,
Spain and France, the countries of the whole European Union, the
Soviet and the Western camps illustrate the existence of instability
in the formation of coalitions during a significant period of their
history.

The mapping from history to the natural model allows us to reproduce
and provide an explanation of these instabilities and the cycling of
coalitions. It then enable us to analyze the conditions necessary to
the system's stability.

Overall, achieving optimal or stable coalitions requires a
significant computation power and a complete availability of the
information. All types of rationality limitations impact the
stabilization and optimization processes.

It can be seen from our study of the model that, as the system's
complexity grows along with the connexity and the size, the stable
or the optimal state become harder to achieve. To overcome those
difficulties, the large systems (such as the United States of
America) tend to subdivide themselves into small weakly connected
sub-systems that manage their internal optimizations and
stabilizations. Large systems that do not follow this course (the
Soviet Union, for example) face difficulties that come from the
complex instabilities communicated to all its parts via the strong
interrelations in the system.

However, a spontaneous maximization that exhibits the bottom-up
dynamics of coalition forming is rare to stabilize. The probability
that the system becomes stable vanishes exponentially with the size.
As a future research, we aim to consider an amalgam of both top-down
and bottom-up dynamics of the coalition forming. The stabilization
by means of an external field could produce such an apposite
amalgam. The external field, while polarizing the interests of the
countries, leads to the emergence of new opposing alliances. The
countries, attaching themselves to one or to the other, find new
interests that unite or separate them based on a pragmatic
motivation instead of the historical concerns. Hereby, the external
field model enables the stabilization among the countries while
keeping the short range character of the interactions between them.
This insight will be investigated in a forthcoming paper.


\begin{thebibliography}{}
\bibitem[Binder \& Young(1986)]{SGM}Binder, K. \& Young, A.P. (1986). Spin-Glasses: experimental facts, theoretical concepts, and open
questions, \emph{Review of Modern Physics}, 58, 801--911

\bibitem[Toulouse (1977)]{GT}Toulouse, G.
  (1977). \emph{Theory of the frustration effect in
Spin Glasses: I}, Comm. on Physics, 2

\bibitem[Galam \& Moscovici(1991)]{MG}Galam, S. \& Moscovici, S. (1991). Towards A Theory Of Collective Phenomena: Concensus And Attitude
Changes In Groups, \emph{European Journal Of Social Psychology}, 21,
49--74

\bibitem[Axelrod \& Bennett (1998)]{Axel} Axelrod, R. \& Bennett, D.S. (1993). A landscape theory of
aggregation, \emph{British Journal Political Sciences}, 23, 211--233

\bibitem[Florian \& Galam (2000)]{Flo} Florian, R. \& Galam, S. (1993). Optimizing
conflicts in the formation of strategic alliances, \emph{Eur. Phys.
J. B }, 16, 189--194


\bibitem[Galam (1998)]{GC} Galam, S. (1998). Comment on A landscape theory of aggregation,
\emph{British Journal Political Sciences}, 28, 411--412


\bibitem[Galam (2002)]{SPC} Galam S. (2002). Spontaneous Coalition Forming. Why Some Are Stable?,
\emph{Springer-Verlag Berlin Heidelberg 2002, Springer-Verlag Berlin
Heidelberg 2002, S. Bandini, B. Chopard, and M. Tomassini
(Eds.):ACRI 2002, LNCS 2493}, 1--9


\bibitem[Galam (1996)]{FVS}  Galam S. (1996).
Fragmentation Versus Stability In Bimodal Coalitions,
\emph{Physica}, A, 230, 174--188


\bibitem[Gerardo \& Samaniego-Steta \& del Castillo-Mussot \& Vazquez (2007)]{TBISCF}
Gerardo G. N. \& Samaniego-Steta F. \& del Castillo-Mussot M. \&
 Vazquez (2007). Three-body interactions in sociophysics and their role in coalition forming,
 \emph{Physica}, A, 379, 226--234


\bibitem[Vinogradova (2012)]{GAVDP} Vinogradova G. (2012).
Correction of Dynamical Network's Viability by Decentralization by
Price, \emph{Journal of Complex Systems}, 20, 1, 37--55

\bibitem[Van Hemmen(1982)]{CSGM} Van Hemmen J.L. (1982).
Classical Spin-Glass Model, \emph{Physical Review Letters}, 49, 6,


\bibitem[Tim Hatamian (2005)]{APIM} Tim Hatamian G. (2005).
On alliance prediction by energy minimization, neutrality and
separation of players, \emph{arxiv.org/pdf/physics/0507017}

\bibitem[Matthews (2000)]{SDO} Matthews R. (2000).
A Spin Glass model of decisions in organizations, \emph{Business
Research Yearbook, G. Biberman, A. Alkhafaji (eds), Saline,
Michigan: McNaughton and Gunn}, 7, 6

\bibitem[Galam \& Gefen \& Shapir(1982)]{SYY} Galam S. \& Gefen Y. \& Shapir Y. (1982).
Sociophysics: A new approach of sociological collective behavior,
\emph{British Journal Political Sciences}, 9, 1–-13.

\bibitem[Antal \& Krapivsky \& Redner(2006)]{SBN} Antal T. \& Krapivsky P.L. \& Redner S. (2006).
Sociophysics: A new approach of sociological collective behavior,
\emph{Physica D}, 224, 130–136.

\bibitem[Acemoglu \& Egorov \& Sonin (2006)]{DSC} Acemoglu D. \& Egorov G. \& Sonin K. (2006).
Dynamics and Stability of Constitutions, Coalitions, and Clubs,
\emph{American Economic Review, American Economic Association},
102(4), 46-76.

\end{thebibliography}
\end{document}